\documentclass[twocolumn]{svjour3}          % twocolumn
\smartqed  % flush right qed marks, e.g. at end of proof
\usepackage{graphicx}
\usepackage{amsmath}
\usepackage{mathptmx}  
\usepackage[square,numbers]{natbib}
\begin{document}

\title{
A statistical approach to
crowdsourced smartphone-based earthquake early warning systems}


\titlerunning{Crowdsourced earthquake early warning systems}

\author{Francesco Finazzi \and Alessandro Fass\`{o}}


\institute{Francesco Finazzi \at
           Dept. of Management, Economics and Quantitative Methods, University of Bergamo \\
           via dei Caniana, 2 - 24127 Bergamo, Italy\\
           \email{francesco.finazzi@unibg.it}
           \and
           Alessandro Fass\`{o} \at
           Dept. of Management, Information and Production Engineering, University of Bergamo \\
					 via Marconi, 5 - 24044 Dalmine, Italy 
					\email{alessandro.fasso@unibg.it}
}

\date{Received: date / Accepted: date}

\maketitle

\begin{abstract}

The Earthquake Network research project implements a crowdsourced earthquake early warning system based on smartphones.
Smartphones, which are made available by the global population, exploit the Internet connection to report a signal to a central server every time a vibration is detected by the on-board accelerometer sensor. 

This paper introduces a statistical approach for the detection of earthquakes from the data coming
from the network of smartphones. The approach allows to handle a dynamic network in which the number of active
nodes constantly changes and where nodes are heterogeneous in terms of sensor sensibility and transmission delay.
Additionally, the approach allows to keep the probability of false alarm under control.

The statistical approach is applied to the data collected by three subnetworks related to the cities
of Santiago de Chile, Iquique (Chile) and Kathmandu (Nepal). The detection capabilities of the approach
are discussed in terms of earthquake magnitude and detection delay.

A simulation study is carried out in order to link the probability of detection and the detection delay to the
behaviour of the network under an earthquake event.

\end{abstract}

\section{Introduction}

Earthquake early warning (EEW) systems are intended to alert population and
first responder agencies when harmful earthquakes strike. These systems are
typically based on networks of stations with sensors measuring seismic and
GPS data in real time \citep{gasparini:2007,Given:2014}. When an earthquake strikes, the
earthquake is detected within few seconds and the population is possibly
notified before strong shaking is experienced.

EEW systems require hundreds of stations located nearby known fault traces.
Construction, maintenance and operation costs can be in the millions of
dollars \citep{Given:2014}. Recently, crowdsourced systems based on network
of smartphones have been exploited in different social and environmental applications \citep{lane:2010}. 
In \citep{overeem:2013}, the temperature measured by smartphones is used to study the so called urban 
heat island effect while the iSPEX project \citep{snik:2014} aims at measuring some properties of
atmospheric aerosols using spectropolarimeters mounted on a large number of
smartphones.

Crowdsourced EEW systems based on smartphones have been theorized and their detection capabilities
have been studied mainly through simulations \citep{Minsonetal:2015}. 
Accelerometric and GPS sensors on-board off-the-shelf smartphones can be exploited to detect
ground shaking and thus possible earthquakes. 
Smartphone-based EEW systems are possibly based on thousands/millions of smartphones with zero
construction costs and near zero maintenance costs. Classic EEW and
smartphone-based EEW systems, however, differ in many respects. First of
all, low-cost sensors on-board smartphones are not specifically designed to
detect earthquakes. Secondly, a crowdsourced EEW system is based on
heterogeneous smartphones from different vendors and thus on different
sensors. Last and most importantly, smartphones of a crowdsourced EEW system
are located where people are, partly reducing the benefits of the EEW system
itself. Nonetheless, an earthquake detected by the smartphones of a city or
town can be notified to the rest of the population not yet reached by the
earthquake.

In this paper, we study the detection capabilities of the crowdsourced EEW
system developed within the Earthquake Network project
(www.earthquakenetwork.it). The detection capabilities are
evaluated in terms of detection rate and detection delay using both real data and simulated data.

The rest of the paper is organized as follows. Section \ref{sec:EN} briefly introduces
the Earthquake Network project while Section \ref{sec:detectors} details the statistical
approach adopted to detect earthquakes from the data coming from the smartphone network.
Section \ref{sec:dataset} introduces the data sets used to demonstrate the detection
capabilities of the statistical approach, which is applied on Section \ref{sec:data_analysis}.
A simulation study on the detection delay is carried out in Section \ref{sec:delay_simulation}
and conclusions are given in Section \ref{sec:conclusions}.

\section{The Earthquake Network project\label{sec:EN}}

Nowadays smartphones are equipped with accelerometric sensors that can be
exploited to measure the ground motion induced by an earthquake
\citep{Dalessandro:2013,Cochran:2012}. The Earthquake Network project aims at
implementing and maintaining a global EEW system based on smartphone
networks, with the smartphones voluntarily made available by the global
population. The Earthquake Network EEW system detects earthquakes in real
time and notifies the population as soon as possible, compatibly with the
latencies of mobile telecommunications technologies.
The project started at the beginning of $2013$
and the system is currently based on a global network of around $70,000$ smartphones. 

The Earthquake Network project is similar in scope to the Quake Catcher
Network (QCN) project \citep{Cochran:2009} and the Community Seismic
Network (CSN) monitoring system \citep{Clayton:2012} though
both QCN and CSN currently rely on Internet-connected computers with
built-in or plugged-in accelerometric sensors. While the scope is similar,
the inherent differences between computers and smartphones imply a different
hardware and software architecture and a different data acquisition
strategy. A key aspect of the Earthquake Network project is that smartphones
have built-in accelerometric sensors while personal computers require an
external device to be installed. It follows that smartphones are part of the
EEW system by simply installing from the Internet the Earthquake Network
smartphone application (\texttt{https://play.google.com/} \newline \texttt{store/apps/details?id=com.finazzi.distquake}).

Although, for reasons of brevity, the hardware and software architecture of the Earthquake Network EEW system are not discussed in details this paper, the essential features of our EEW system are sketched here. 

When the smartphone owner installs the Earthquake Network application, the smartphone becomes a node of the EEW system.
A smartphone enters the \textit{active status} after that it has been connected to a source of power (to avoid battery drainage), that it has been recognized to be in a stationary condition (not in use and not exposed to systematic movements) and that a sensor calibration has been performed. 

An active smartphone constantly monitors its acceleration and it sends an \textit{active signal} to the server every $30$ minutes. Active signals are used to describe the state of the network, in terms of number of active nodes and their spatial distribution.  

When an active smartphone detects a vibration, it sends a \textit{vibration signal} to the server with information on the
geographic position of the smartphone (latitude and longitude). 
If the number of signals received is high with respect to the active smartphones for a certain area, the server issues an earthquake warning to the network.

This paper focuses on the detection algorithm implemented server-side. At the current stage of the project, epicenter and magnitude are not estimated and the earthquake is simply located at the centroid of the area where the earthquake is detected. Epicenter and magnitude estimation techniques for EEW systems are well studied in literature \citep{lancieri:2008,satriano:2011} and their applicability to the Earthquake Network EEW system is still under evaluation.

\section{Statistical EEW\label{sec:detectors}}

This section introduces a strategy which aims at detecting an earthquake as soon as possible by the real-time analysis of the data collected by the network of smartphones and sent to the server. 

In classic EEW systems, an earthquake warning can be issued when one or few seismometers in the network detect the first P-waves generated by the earthquake (see for instance \citealp{cua:2009}). 
In the case of the Earthquake Network EEW system, it is observed that each smartphone in the network
sends, on average, around $30$ vibration signals per day not related to earthquakes.
Indeed, though the application running on the smartphone filters some of these signals, many of them are not discriminable at the smartphone level and, henceforth, they are sent to the server. It follows that the server must be able to detect an earthquake by identifying vibration signals which are related to a real earthquake. 

In the sequel, the term \textit{false signal} will be used to denote a false vibration signal sent
by a smartphone while the term \textit{false alarm} will be used to denote a false earthquake warning issued by the detector. In order to keep the probability of false alarm under control, a statistical strategy is adopted and a statistical algorithm used for detection will be called here a \textit{detector}. 

Since the network of smartphones do not share a common clock, the server time is considered as reference. 
Transmission delays are not measured nor estimated and they are assumed to be negligible with respect to the detection problem. Moreover, a fixed spatial area is considered, such as a city or a small region.

\subsection{Stochastic modeling of vibration signals\label{par:arrival_times}}

Arrival times of the vibration signals are modeled through a stochastic
point process under the hypothesis of no earthquake and deviation from this
hypothesis is tested each time a vibration signal reaches the server.

Let $\left\{ N\left( t\right) ,t>t_{0}\right\} $ be the stochastic point
process giving the probabilistic framework for the observed vibration
signals coming from the network at arrival times $t_{1}<...<t_{T}$. 
In particular, under the no earthquake hypothesis, $N\left( t\right) $ is assumed to be a Poisson process with conditional intensity function 
\begin{equation}
\lambda ^{0}\left(t\right) = \lambda ^{0}\left( t,\mathbf{x}_t\right) 
\nonumber
\end{equation}
where $\mathbf{x}_t$ is a vector of covariates.
According to standard notation, the number of vibration signals
in the interval $I=(a,b]$ is denoted by $N\left( I\right) $ and is known to have Poisson distribution with expectation 
\begin{equation}
E\left[ N\left( I\right) \mid \mathbf{x}_t \right] =
\Lambda \left( I \mid \mathbf{x}_t\right) =\int_{I}\lambda^{0}\left( t\right) dt.
\nonumber
\end{equation}
Notice that, despite the no earthquake hypothesis, $\lambda$ may show relevant variations due to day/night behavior and trends in the network size driven by external factors. In particular 
the number of active smartphones at time $t$ is a relevant quantity.
Although this quantity is difficult to be known exactly in real time, a proxy given by the number of active signals in the last 30 minutes, say $\nu_{t}$, is routinely available on the server with cheap computational cost and, with abuse of language, it will be called number of active smartphones in the sequel. 
As a result a Poissonian GLM is used with 
\begin{equation}
\lambda ^{0}\left( t\right)=\exp \left( \beta _{0}+\beta _{1}\nu_{t}\right)  \label{eq:glm}
\end{equation}%
where $\mathbf{\beta =}\left( \beta _{0},\beta _{1}\right) ^{\prime }$ is
the parameter vector to be estimated on historical no earthquake data. 

\subsection{Continuous time detectors\label{par:continuous_detection}}

As the earthquake waves
propagate at a given speed from the epicenter, the earthquake is not
instantly felt by all the active smartphones in the network. Additionally, random transmission delays are possible. 
For these reasons, the vibration signals in the interval $I_{\varepsilon
}^{t}=(t-\varepsilon ,t]$, for some $\varepsilon >0$, are considered as information related to the same earthquake and their number is denote by $%
N_{\varepsilon }^{t}=N\left( I_{\varepsilon }^{t}\right) $. 

Since $\varepsilon $ is relatively small
(e.g. $30$ $s$) it is not easy to use standard change point detection
techniques \citep{Basseville:1993} which are tailored for
permanent changes and asymptotic theory. Additionally, control chart techniques
based on the Poisson distribution \citep{borror:1998,he:2006,mei:2011} are not
useful here as $N_{\varepsilon }^{t}$ is computed at each vibration signal arrival and, especially
under an earthquake event, the $N_{\varepsilon }^{t}$ are not i.i.d.

Hence two likelihood approaches based on the generalized likelihood ratio (GLR) statistic (see for instance \citep{capizzi:2008}) and on the efficient score, respectively, are considered in the sequel. Although the first one will result in a detector which is too slow to be computed in real time, it is presented here because it is a useful introduction to the score detector.

We begin with the well known
log-likelihood of the signals in the interval $I_{\varepsilon }^{t}$ \citep{Snyder:2012} which is given by%
\begin{equation}
\log L\left( \lambda |t,\varepsilon \right) =\sum_{t_{j}\in I_{\varepsilon
}^{t}}\log \lambda \left( t_{j}\right) -\Lambda \left( I_{\varepsilon
}^{t}\right)  \label{eq:log-lik}
\end{equation}%
where $t_{j}$ are the arrival times of the vibration signals in $I_{\varepsilon }^{t}$.

Now suppose that, under a seismic event, the process intensity has a peak given by%
\begin{equation}
\lambda\left( t\right) =\lambda ^{0}\left( t\right) +\frac{\Delta }{%
\varepsilon }  \label{eq:lambda_1}
\nonumber
\end{equation}%
with $\Delta >0$ for $t\in I_{\varepsilon }^{t}$ and $\Delta =0$ otherwise.
The log-likelihood in equation ($\ref{eq:log-lik}$) has the following form%
\begin{equation}
\log L\left( \Delta \right) =\sum_{t_{j}\in I_{\varepsilon }^{t}}\log \left(
\lambda ^{0}\left( t_{j}\right) +\frac{\Delta }{\varepsilon }\right)
-\Lambda ^{0}\left( I_{\varepsilon }^{t}\right) -\Delta.
\nonumber
\end{equation}%
For fixed $\varepsilon $, the GLR statistic is given by%
\begin{align}
GLR\left( \varepsilon ,t\right) & =\log \frac{L\left( \hat{\Delta}%
_{\varepsilon }^{t}\right) }{L\left( 0\right) } 
\nonumber
\\
& =\sum_{t_{j}\in I_{\varepsilon }^{t}}\log \left( 1+\frac{\hat{\Delta}%
_{\varepsilon }^{t}}{\varepsilon \lambda ^{0}\left( t_{j}\right) }\right) -%
\hat{\Delta}_{\varepsilon }^{t}
\nonumber
\end{align}%
where 
\begin{equation*}
\hat{\Delta}_{\varepsilon }^{t}=\max \left( 0,\underset{\Delta }{\arg \max }%
L\left( \Delta \right) \right)
\end{equation*}%
and where $\underset{\Delta }{\arg \max }L\left( \Delta \right) $ is given
by the solution of the following likelihood equation%
\begin{equation}
\sum_{t_{j}\in I_{\varepsilon }^{t}}\frac{1}{\varepsilon \lambda ^{0}\left(
t_{j}\right) +\Delta }-1=0.  \label{eq:likelihood_equation}
\nonumber
\end{equation}%
This can be solved numerically using as initial value the method of moment estimate of $%
\Delta $, which is easily seen to be given by $\tilde{\Delta}=N_{\varepsilon
}^{t}-\Lambda ^{0}\left( I_{\varepsilon }^{t}\right) $. The above GLR
depends on the detection interval size $\varepsilon $ which is arbitrary.
Hence, extending \citep{Basseville:1993}, a GLR detector gives an earthquake warning if
\begin{equation}
\sup_{\varepsilon >0}GLR\left( \varepsilon ,t\right) >h  \label{eq:GLR}
\nonumber
\end{equation}%
for some threshold $h$ and $\lambda ^{0}\left( t\right)$ is computed using equation (\ref{eq:glm}).

The second likelihood approach is based on the efficient score which is
given by%
\begin{equation}
S\left( \varepsilon ,t\right) =\left. \frac{\partial }{\partial \Delta }\log
L\left( \Delta \right) \right\vert _{\Delta =0}=\sum_{t_{j}\in
I_{\varepsilon }^{t}}\frac{1}{\varepsilon \lambda ^{0}\left( t_{j}\right) }-1
\label{eq:score_time_variant}
\nonumber
\end{equation}%
and the score detector gives an earthquake warning if%
\begin{equation}
\sup_{\varepsilon >0}S\left( \varepsilon ,t\right) >h  \label{eq:Score_detector}
\end{equation}%
for some threshold $h$. 

For small $\epsilon$, the intensity $\lambda ^{0}$ can be assumed approximately constant in $I_{\varepsilon }^{t}$ and considering the fast dynamics of the earthquake, an approximate score detector is given by
\begin{equation}
S\left( \varepsilon ,t\right) \cong \frac{N_{\varepsilon }^{t}}{\varepsilon
\lambda^0(t)}-1  \label{eq:score_time_invariant} > h
%\label{eq:Approx_Score_detector}
\end{equation}%
for some $h$, which is quite faster to be computed with respect to detector given by (\ref{eq:Score_detector}).

\subsection{Threshold modeling\label{sec:thresholds}}

A critical aspect for the detectors defined above is the choice of the
threshold $h$. 
Indeed, there is a trade-off between the probability of false alarm 
$\alpha = P(S(\epsilon,t)>h \mid \Delta=0)$ and the probability of missed detection of an earthquake, namely $P(S(\epsilon,t)<h \mid \Delta=\Delta^*)$ for a certain  $\Delta^*>0$.
A low value of $h$ implies a higher detection probability but also a higher probability of false alarm. 
Additionally, a higher value of $h$ implies a higher detection
delay when a real earthquake is striking.
Since an uncontrolled and possibly large number of false alarms makes  the warning system ineffective and at hand lead people to abandon the network, we give higher priority at controlling the probability of false alarm $\alpha$. 

It is than a natural choice to take $h$ as a quantile of the
distribution of the score $S( \epsilon,t)$ under the null hypothesis of no earthquake, corresponding to  a very small $\alpha$ or a very large time between false alarms which is approximately $1/\alpha$.
Unfortunately the exact distribution of $S( \epsilon,t)$ is not readily available and the asymptotic normality of the score statistic cannot be used here because $\epsilon$ is small and the Gaussian approximation is not satisfactory, especially in the far right tail of the detector distribution.
The tail distribution is then estimated directly on data, thus giving a robust approach against violation of the parametric assumptions of the previous section.

To do this, firstly, a long sample is extracted from the dataset under no earthquake conditions.
Secondly, the right tail of the empirical distribution above the $p_{0}$ quantile (say $p_{0}=0.99$) is considered and modeled through a
generalized Pareto distribution. 
In order to have an average of one false alarms over the period $\Delta T$ (e.g. one false alarm per year), 
we set $\alpha = \overline{\Delta t}/\Delta T$ 
where $\overline{\Delta t}$ is the observed
mean time between (false) vibration signals.
Finally, $h$ is given by the $p_1$ quantile of the generalized Pareto distribution, with $p_1=1-\frac{\alpha}{1-p_{0}}$, 

\section{Earthquake data\label{sec:dataset}}

The Earthquake Network project started on January 1st, 2013. Since then, the
Android application has been downloaded more than $650,000$ times
and the network has grown up to around $70,000$ users globally. As
participation to the network is voluntary, the number of users in the
network changes continuously as well as their spatial distribution. In
general, strong earthquakes felt by the population induce a large number of
downloads and the spontaneous growth of new subnetworks. In many cases,
subnetworks coincide with cities where users are clustered. Subnetworks are
characterized by their own life cycle and they may disappear if users lose
interest.

This paper considers the three subnetworks of Santiago de Chile (Chile), Iquique (Chile)\ and Kathmandu (Nepal), in year 2015 as specified in Table \ref{tab:subnetworks}.
The subnetworks of Santiago de Chile and Iquique are quite stable in time, with the former bigger than the latter. Conversely, the Kathmandu subnetwork has grown rapidly after the $7.8$ magnitude earthquake that hit Nepal on April 25, 2015. 

\begin{figure}
  \includegraphics[width=0.48\textwidth]{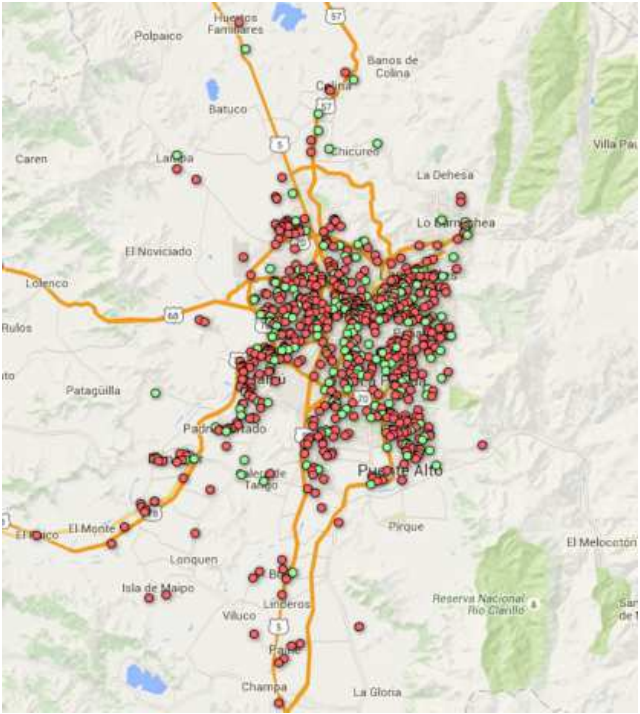}
\caption{Santiago de Chile subnetwork at 6:30 (local time) of a typical working day. Green dots are the active smartphone while the red dots are smartphone connected to Internet but not active at that time.}
\label{fig:santiago_subnetwork}     
\end{figure}

Figure \ref{fig:santiago_subnetwork} shows
the Santiago de Chile subnetwork in the early morning of a working day. 
Note that the spatial pattern of the smartphones reflects the population distribution within the
subnetwork. Also note that only a fraction of the smartphones is active (green dots).

In order to demonstrate the functioning of the detector $S\left( \varepsilon
,t\right) $ above introduced, data coming from each subnetwork have been
collected over the time frame reported in Table \ref{tab:subnetworks}. 
Statistics related to the active smartphones, given in
the same table, refer to the number of smartphones which are active at any
given time during the day. This number is significantly lower than the
number of users which take part in the subnetwork as, at any given time, a
fraction of smartphones is switched off and another fraction is not active.
The number of active smartphones largely changes during the day and it is
usually higher at night (when people charge their smartphones) and lower at
daytime. A weekend effect is also observed, with the average number of
active smartphone which is lower during the weekend.

\begin{table*}[tbp]
\centering
\caption{Subnetwork details and statistics of active smartphones. Symbols $\tilde{\nu}_{5}$, $\bar{\nu}$ and $\tilde{\nu}_{95}$ denote, respectively, 5th percentile, average and 95th percentile of the number of active smartphones.}
\label{tab:subnetworks}%
\begin{tabular}{ccccccc}
\hline
&  &  &  & \multicolumn{3}{c}{Active smartphones} \\ 
Subnetwork & Diameter & Population & Time frame& $\tilde{\nu}_{5}$ & $\bar{\nu}$ & 
$\tilde{\nu}_{95}$ \\ \hline
\multicolumn{1}{l}{Santiago (Chile)} & \multicolumn{1}{r}{$40$ $km$} & 
\multicolumn{1}{r}{$6,300,000$} & \multicolumn{1}{l}{Jan
7, 2015 - Apr 9, 2015} & \multicolumn{1}{r}{$51$} & \multicolumn{1}{r}{$183$} & 
\multicolumn{1}{r}{$416$} \\ 
\multicolumn{1}{l}{Iquique (Chile)} & \multicolumn{1}{r}{$15$ $km$} & 
\multicolumn{1}{r}{$182,000$} & \multicolumn{1}{l}{Jan 7, 2015 - Apr 9, 2015} & 
\multicolumn{1}{r}{$29$} & \multicolumn{1}{r}{$78$} & \multicolumn{1}{r}{$165
$} \\ 
\multicolumn{1}{l}{Kathmandu (Nepal)} & \multicolumn{1}{r}{$30$ $km$} & 
\multicolumn{1}{r}{$1,000,000$} & \multicolumn{1}{l}{Apr
25, 2015 - May 15, 2015} & \multicolumn{1}{r}{$15$} & \multicolumn{1}{r}{$38$} & 
\multicolumn{1}{r}{$70$} \\ \hline
\end{tabular}%
\end{table*}

\section{Data analysis \label{sec:data_analysis}}

Data collected from each subnetwork consist in the time-ordered list, $%
\mathcal{L}$ say, of the timing of each vibration signal received by the server, together with smartphone georeferencing (latitude and longitude) and the estimated number of active smartphones $\nu_{t}$ in the subnetwork. 
% Latitude and longitude are not currently used and they simply allow to discriminate across subnetworks.
% non è vero sono usate per definire L^0

As the statistical distribution of the detector $S\left( \varepsilon
,t\right) $ is derived under the hypothesis of no earthquake, it is
necessary to obtain the list $\mathcal{L}^{0}$ of all the vibration signals
not induced by a real earthquake (false signals). This is done considering
the catalog of the European Mediterranean Seismological Centre (EMSC) which
provides information on the earthquakes detected globally. Earthquakes
occurred within a radius of $1000$ $km$ from each subnetwork, and likely
felt, are firstly identified and then used to derive $\mathcal{L}^{0}$
removing from $\mathcal{L}$ all the signals received from the beginning of each earthquake up to $5$ minutes later. Indeed, even if an earthquake duration is in the range of seconds, after a seismic event, the network may experience a sudden transient related to phone lines crowding, switch off and other outlying behaviors.

The choice of the window size $\varepsilon $ may influence the detection probability and the probability of false alarm. In this work $\epsilon=30$ $s$ is used and its choice is justified in Section \ref{sec:delay_simulation} by simulation results.

\subsection{Parameter estimation}

The vibration signals in the list $\mathcal{L}^{0}$ are used to estimate the parameters of $\lambda^{0}(t)=\exp \left( \beta _{0}+\beta _{1}\nu_{t}\right) $ which is needed to
compute $S\left( \varepsilon ,t\right) $ as in equation 
(\ref{eq:score_time_invariant}). 
Table \ref{tab:threshold_estimation} reports the estimates of $\beta _{0}$ and $\beta_{1}$, which are obtained by the maximum likelihood method and have standard deviations smaller than $9 \times 10^{-3}$ and $2 \times 10^{-4}$, respectively, for all the subnetworks. 

\begin{figure}
  \includegraphics[width=0.48\textwidth]{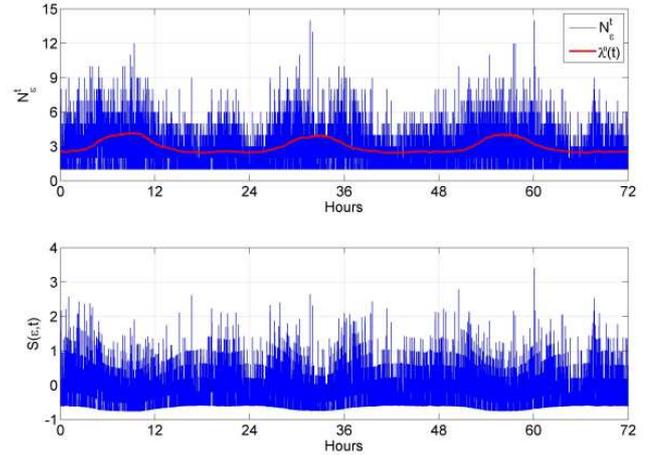}
\caption{Graphs of $N_{\varepsilon }^{t}$, $\lambda^0(t)$ and $S\left( \varepsilon ,t\right)$ evaluated
for the Santiago de Chile subnetwork over $3$ days starting from February 25, 2015 00:00 (UTC).}
\label{fig:santiago_score}     
\end{figure}

Considering $\varepsilon=30$ $s$, equation (\ref{eq:score_time_invariant}) is used to compute $%
S\left( \varepsilon ,t\right) $ for each signal in $\mathcal{L}^{0}$. As an example, Figure \ref{fig:santiago_score} shows the graphs of $N_{\epsilon }^{t}$, $\lambda^0(t)$ and $S(\epsilon ,t)$ for the Santiago de Chile subnetwork over a period of 3 days and for around $18,600$ vibration signals received. 
Note that $N_{\varepsilon }^{t}$ reflects the
daily cycle of the number of active smartphones and that the server received
up to $14$ vibration signals in less than $30$ $s$ even in the absence of earthquakes. 
This shows that each smartphone, taken individually, is not a reliable
seismometer and a statistical approach, which is able to discriminate
real earthquakes, is useful.

Following the procedure detailed in Section \ref{sec:thresholds} and
considering $p_{0}=0.99$, the empirical distribution of $S\left( \varepsilon
,t\right) $ is analysed to derive a threshold $h$ corresponding to a mean time between false alarms of approximately one year. Table \ref{tab:threshold_estimation} reports the intermediate
results and the estimated $h$.

\begin{table*}[tbp] 
\centering
\caption{GLM parameter and threshold estimation. Legend: $\left\vert \mathcal{L}^{0}\right\vert$: number of false signals observed over
the time frame specified in Table \ref{tab:subnetworks}; $\hat{\beta}_{0}$ and $\hat{\beta}_{1}$: estimated parameters of the GLM;
$\overline{\Delta t}$ $\left( s\right) $: average inter-arrival time; $p_{1}$: percentile of the Pareto distribution; $h$: estimated threshold.}
\label{tab:threshold_estimation}%
\begin{tabular}{ccccccc}
\hline
&  & \multicolumn{2}{c}{GLM} & \multicolumn{3}{c}{Threshold estimation} \\ 
Subnetwork & $\left\vert \mathcal{L}^{0}\right\vert $ & $\hat{\beta}_{0}$ & $%
\hat{\beta}_{1}$ & $\overline{\Delta t}$ $\left( s\right) $ & $p_{1}$ & $h$
\\ \hline
\multicolumn{1}{l}{Santiago de Chile} & \multicolumn{1}{r}{$442,000$%
} & $0.7694$ & $0.0016$ & \multicolumn{1}{r}{$18.0$} & $0.99994$ & 
\multicolumn{1}{r}{$6.42$} \\ 
\multicolumn{1}{l}{Iquique} & \multicolumn{1}{r}{$208,000$} & $%
0.4111$ & $0.0027$ & \multicolumn{1}{r}{$38.2$} & $0.99988$ & 
\multicolumn{1}{r}{$4.21$} \\ 
\multicolumn{1}{l}{Kathmandu} & \multicolumn{1}{r}{$19,400$} & $%
0.1190$ & $0.0068$ & \multicolumn{1}{r}{$88.6$} & $0.99972$ & 
\multicolumn{1}{r}{$4.16$} \\ \hline
\end{tabular}%
\end{table*}%

\subsection{Earthquake detection}

With the threshold $h$ available, the detector $S\left( \varepsilon
,t\right) $ is used to detect an earthquake every time $S\left( \varepsilon
,t\right) $ exceeds $h$. Typically, the detector works in an on-line manner,
that is, $S\left( \varepsilon ,t\right) $ is computed every time the server
receives a vibration signal and an earthquake warning is issued when $%
S\left( \varepsilon ,t\right) >h$. Here, the detector is applied off-line to
the list $\mathcal{L}$ in order to show its detection capabilities on past
earthquakes.

Note that the detector is not guaranteed to detect all the earthquakes that
was felt in a given subnetwork. If the number of active smartphones is too
low and/or the earthquake is very mild, then the earthquake may not be
recognized from the ``background noise".

Tables \ref{tab:detected} reports, for each subnetwork, the information on
the earthquakes which are detected over the time frames specified in Table \ref{tab:subnetworks}. Earthquake time, latitude, longitude, depth and
magnitude are obtained from the above mentioned EMSC catalog while the
distance is computed from the epicenter to the centre of the subnetwork. The
detection time $t^{\ast }$ refers to the time at which $S\left( \varepsilon
,t\right) $ exceeds $h$ while the delay is computed as the difference
between $t^{\ast }$ and the first vibration signal received by the server
and related to the earthquake. The table also reports the number of active
smartphones at $t^{\ast }$, the number of vibration signals induced by the
earthquake and the fraction of active smartphones that sent a vibration
signal.

As different earthquakes are characterized by different parameters (mainly
magnitude, depth and distance) it is not easy to derive final conclusions on
the behaviour of the subnetworks under an earthquake event. Nonetheless, some general comments can be made. 
First of all, mild earthquakes (say below magnitude $4$) are not detected. 
The accelerometer on-board the smartphone may not be sensible enough to detect a vibration or the earthquake is detected by a very small number of smartphones. 
Secondly, the fraction of active smartphones reporting a vibration signal is, on average, $0.31$ and it is rarely higher even in the case of strong earthquakes. 
This is because, by chance, smartphones may be located over soft and vibration absorbing surfaces which reduce the acceleration induced by an earthquake, or the Internet connection may not be available at the time the earthquake is felt. 
Finally, the detection delay (which does not include the delay due to the distance from the epicenter) ranges from $2$ to $17$ $s$. This delay is influenced by the temporal spread of the vibration signals which are sent to the server; the lower the spread the lower the detection delay.

Note that two of the earthquakes detected by the subnetworks are actually false alarms. False alarms
are likely related to the behaviour of the users with the smartphone application installed. Indeed,
the application allows to receive many kind of notifications (e.g. the notifications related to
the earthquakes detected by the national and international seismic networks). Since these
notifications are received by many users within a short time frame, the same users may
induce multiple vibrations (for instance by picking up the smartphone) which are sent to the 
server and which are identified as an earthquake. This problem will be solved with future released of the
smartphone application, allowing to reduce the occurrence of ``clusters" of false vibration signal, to lower
the threshold $h$ and thus to lower the detection delay.

\begin{table*}[tbp]
\centering
\caption{Real earthquakes and false alarms signaled by $S(\epsilon,t)$. Legend: Earthquake time: UTC of the earthquake from EMSC catalog; Lat.,Lon.: epicenter location; Mag.: earthquake magnitude; Dist.: distance between epicenter and subnetwork centre; Detection time: UTC of detection; Delay: detection delay wrt first vibration signal related to the earthquake; 
$\nu_{t}$: number of active smartphones at detection time;
$\tilde{N}$: total number of vibration signals received by the server during the earthquake; $f$: fraction
of active smartphones that reported the earthquake.}
\label{tab:detected}
\begin{tabular}{ccccccccccc}
\hline
\multicolumn{11}{c}{\textbf{Santiago de Chile subnetwork}} \\ 
Earthquake time & Lat. & Lon. & Depth & Mag. & Dist. & Detection time & Delay
& $\nu_{t}$ & $\tilde{N}$ & $f$ \\ \hline
\multicolumn{1}{r}{15/01/2015 05:19:45} & \multicolumn{1}{r}{$-33.61%
{{}^\circ}%
$} & \multicolumn{1}{r}{$-71.22%
{{}^\circ}%
$} & \multicolumn{1}{r}{$80$ $km$} & \multicolumn{1}{r}{$mb$ $4.6$} & 
\multicolumn{1}{r}{$56$ $km$} & \multicolumn{1}{r}{15/01/2015 05:20:09} & 
\multicolumn{1}{r}{$6$ $s$} & \multicolumn{1}{r}{$147$} & \multicolumn{1}{r}{%
$73$} & \multicolumn{1}{r}{$0.50$} \\ 
\multicolumn{1}{r}{25/01/2015 08:47:04} & \multicolumn{1}{r}{$-34.72%
{{}^\circ}%
$} & \multicolumn{1}{r}{$-71.67%
{{}^\circ}%
$} & \multicolumn{1}{r}{$40$ $km$} & \multicolumn{1}{r}{$mb$ $4.7$} & 
\multicolumn{1}{r}{$168$ $km$} & \multicolumn{1}{r}{25/01/2015 08:47:50} & 
\multicolumn{1}{r}{$11$ $s$} & \multicolumn{1}{r}{$151$} & 
\multicolumn{1}{r}{$50$} & \multicolumn{1}{r}{$0.33$} \\ 
\multicolumn{1}{l}{False alarm} & \multicolumn{1}{r}{} & \multicolumn{1}{r}{}
& \multicolumn{1}{r}{} & \multicolumn{1}{r}{} & \multicolumn{1}{r}{} & 
\multicolumn{1}{r}{02/02/2015 10:51:39} & \multicolumn{1}{r}{} & 
\multicolumn{1}{r}{$164$} & \multicolumn{1}{r}{$21$} & \multicolumn{1}{r}{$%
0.13$} \\ 
\multicolumn{1}{r}{17/02/2015 14:35:55} & \multicolumn{1}{r}{$-32.33%
{{}^\circ}%
$} & \multicolumn{1}{r}{$-70.74%
{{}^\circ}%
$} & \multicolumn{1}{r}{$94$ $km$} & \multicolumn{1}{r}{$Mw$ $5.4$} & 
\multicolumn{1}{r}{$127$ $km$} & \multicolumn{1}{r}{17/02/2015 14.36.37} & 
\multicolumn{1}{r}{$17$ $s$} & \multicolumn{1}{r}{$71$} & \multicolumn{1}{r}{%
$24$} & \multicolumn{1}{r}{$0.34$} \\ 
\multicolumn{1}{r}{24/02/2015 05:14:02} & \multicolumn{1}{r}{$-32.63%
{{}^\circ}%
$} & \multicolumn{1}{r}{$-71.71%
{{}^\circ}%
$} & \multicolumn{1}{r}{$60$ $km$} & \multicolumn{1}{r}{$mb$ $4.9$} & 
\multicolumn{1}{r}{$136$ $km$} & \multicolumn{1}{r}{24/02/2015 05:14:45} & 
\multicolumn{1}{r}{$10$ $s$} & \multicolumn{1}{r}{$303$} & 
\multicolumn{1}{r}{$101$} & \multicolumn{1}{r}{$0.33$} \\ 
\multicolumn{1}{r}{01/04/2015 15:54:14} & \multicolumn{1}{r}{$-33.74%
{{}^\circ}%
$} & \multicolumn{1}{r}{$-71.02%
{{}^\circ}%
$} & \multicolumn{1}{r}{$67$ $km$} & \multicolumn{1}{r}{$ML$ $4.0$} & 
\multicolumn{1}{r}{$46$ $km$} & \multicolumn{1}{r}{01/04/2015 15:54:43} & 
\multicolumn{1}{r}{$12$ $s$} & \multicolumn{1}{r}{$93$} & \multicolumn{1}{r}{%
$23$} & \multicolumn{1}{r}{$0.25$} \\ 
\multicolumn{1}{r}{} & \multicolumn{1}{r}{} & \multicolumn{1}{r}{} & 
\multicolumn{1}{r}{} & \multicolumn{1}{r}{} & \multicolumn{1}{r}{} & 
\multicolumn{1}{r}{} & \multicolumn{1}{r}{} & \multicolumn{1}{r}{} & 
\multicolumn{1}{r}{} & \multicolumn{1}{r}{} \\ \hline
\multicolumn{11}{c}{\textbf{Iquique subnetwork}} \\ 
Earthquake time & Lat. & Lon. & Depth & Mag. & Dist. & Detection time & Delay
& $\nu_{t}$ & $\tilde{N}$ & $f$ \\ \hline
\multicolumn{1}{r}{09/01/2015 11:48:28} & \multicolumn{1}{r}{$-20.43%
{{}^\circ}%
$} & \multicolumn{1}{r}{$-68.94%
{{}^\circ}%
$} & \multicolumn{1}{r}{$109$ $km$} & \multicolumn{1}{r}{$Mw$ $4.8$} & 
\multicolumn{1}{r}{$128$ $km$} & \multicolumn{1}{r}{09/01/2015 11:49:12} & 
\multicolumn{1}{r}{$7$ $s$} & \multicolumn{1}{r}{$70$} & \multicolumn{1}{r}{$%
24$} & \multicolumn{1}{r}{$0.34$} \\ 
\multicolumn{1}{r}{24/02/2015 05:13:50} & \multicolumn{1}{r}{$-22.70%
{{}^\circ}%
$} & \multicolumn{1}{r}{$-66.68%
{{}^\circ}%
$} & \multicolumn{1}{r}{$182$ $km$} & \multicolumn{1}{r}{$mb$ $5.3$} & 
\multicolumn{1}{r}{$452$ $km$} & \multicolumn{1}{r}{24/02/2015 05:14:51} & 
\multicolumn{1}{r}{$7$ $s$} & \multicolumn{1}{r}{$119$} & \multicolumn{1}{r}{%
$12$} & \multicolumn{1}{r}{$0.10$} \\ 
\multicolumn{1}{r}{03/03/2015 12:45:18} & \multicolumn{1}{r}{$-20.39%
{{}^\circ}%
$} & \multicolumn{1}{r}{$-69.03%
{{}^\circ}%
$} & \multicolumn{1}{r}{$104$ $km$} & \multicolumn{1}{r}{$Mw$ $5.1$} & 
\multicolumn{1}{r}{$118$ $km$} & \multicolumn{1}{r}{03/03/2015 12:45:49} & 
\multicolumn{1}{r}{$7$ $s$} & \multicolumn{1}{r}{$47$} & \multicolumn{1}{r}{$%
9$} & \multicolumn{1}{r}{$0.19$} \\ 
\multicolumn{1}{r}{09/03/2015 03:22:20} & \multicolumn{1}{r}{$-19.70%
{{}^\circ}%
$} & \multicolumn{1}{r}{$-69.33%
{{}^\circ}%
$} & \multicolumn{1}{r}{$91$ $km$} & \multicolumn{1}{r}{$mb$ $4.7$} & 
\multicolumn{1}{r}{$102$ $km$} & \multicolumn{1}{r}{09/03/2015 03:22:59} & 
\multicolumn{1}{r}{$15$ $s$} & \multicolumn{1}{r}{$102$} & 
\multicolumn{1}{r}{$17$} & \multicolumn{1}{r}{$0.17$} \\ 
\multicolumn{1}{l}{False alarm} & \multicolumn{1}{r}{} & \multicolumn{1}{r}{}
& \multicolumn{1}{r}{} & \multicolumn{1}{r}{} & \multicolumn{1}{r}{} & 
\multicolumn{1}{r}{14/03/2015 06.24.32} & \multicolumn{1}{r}{} & 
\multicolumn{1}{r}{$140$} & \multicolumn{1}{r}{$12$} & \multicolumn{1}{r}{$%
0.09$} \\ 
\multicolumn{1}{r}{23/03/2015 04:51:38} & \multicolumn{1}{r}{$-18.46%
{{}^\circ}%
$} & \multicolumn{1}{r}{$-69.17%
{{}^\circ}%
$} & \multicolumn{1}{r}{$132$ $km$} & \multicolumn{1}{r}{$Mw$ $6.4$} & 
\multicolumn{1}{r}{$220$ $km$} & \multicolumn{1}{r}{23/03/2015 04:52:16} & 
\multicolumn{1}{r}{$3$ $s$} & \multicolumn{1}{r}{$124$} & \multicolumn{1}{r}{%
$51$} & \multicolumn{1}{r}{$0.41$} \\ 
\multicolumn{1}{r}{} & \multicolumn{1}{r}{} & \multicolumn{1}{r}{} & 
\multicolumn{1}{r}{} & \multicolumn{1}{r}{} & \multicolumn{1}{r}{} & 
\multicolumn{1}{r}{} & \multicolumn{1}{r}{} & \multicolumn{1}{r}{} & 
\multicolumn{1}{r}{} & \multicolumn{1}{r}{} \\ \hline
\multicolumn{11}{c}{\textbf{Kathmandu subnetwork}} \\ 
Earthquake time & Lat. & Lon. & Depth & Mag. & Dist. & Detection time & Delay
& $\nu_{t}$ & $\tilde{N}$ & $f$ \\ \hline
\multicolumn{1}{r}{12/05/2015 07:05:19} & \multicolumn{1}{r}{$27.89$} & 
\multicolumn{1}{r}{$86.17$} & \multicolumn{1}{r}{$10$ $km$} & 
\multicolumn{1}{r}{$7.3$ $Mw$} & \multicolumn{1}{r}{$85$ $km$} & 
\multicolumn{1}{r}{12/05/2015 07:05:42} & \multicolumn{1}{r}{$4$ $s$} & 
\multicolumn{1}{r}{$32$} & \multicolumn{1}{r}{$11$} & \multicolumn{1}{r}{$%
0.34$} \\ 
\multicolumn{1}{r}{12/05/2015 20:22:15} & \multicolumn{1}{r}{$27.57$} & 
\multicolumn{1}{r}{$85.06$} & \multicolumn{1}{r}{$10$ $km$} & 
\multicolumn{1}{r}{$4.5$ $mb$} & \multicolumn{1}{r}{$30$ $km$} & 
\multicolumn{1}{r}{12/05/2015 20:22:21} & \multicolumn{1}{r}{$2$ $s$} & 
\multicolumn{1}{r}{$42$} & \multicolumn{1}{r}{$23$} & \multicolumn{1}{r}{$%
0.55$} \\ 
\multicolumn{1}{r}{15/05/2015 01:42:43} & \multicolumn{1}{r}{$28.09$} & 
\multicolumn{1}{r}{$84.9$} & \multicolumn{1}{r}{$10$ $km$} & 
\multicolumn{1}{r}{$4.9$ $mb$} & \multicolumn{1}{r}{$61$ $km$} & 
\multicolumn{1}{r}{15/05/2015 01:43:06} & \multicolumn{1}{r}{$5$ $s$} & 
\multicolumn{1}{r}{$87$} & \multicolumn{1}{r}{$14$} & \multicolumn{1}{r}{$%
0.16$} \\ \hline
\end{tabular}%
\end{table*}%

\begin{figure}
  \includegraphics[width=0.48\textwidth]{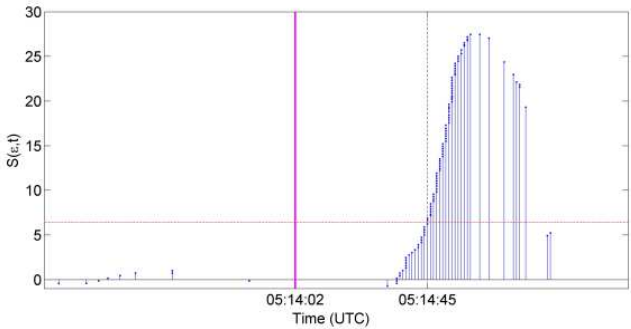}
\caption{Graph of $S\left(\varepsilon,t\right)$ before and after the earthquake detected
by the Santiago de Chile subnetwork on February 24, 2015 05:14:55 (UTC). Legend: vertical bars: value of $S\left(\varepsilon,t\right)$ computed at each vibration signal; vertical solid line: time of the earthquake at the epicenter; vertical dashed line: detection time; horizontal dashed line: threshold $h$.}
\label{fig:detection}     
\end{figure}

As an example, Figure \ref{fig:detection} shows the behavior of 
$S(\epsilon ,t) $ for an earthquake that affected the subnetwork of
Santiago de Chile. The earthquake struck on February 24, 2015 at 05:14:02
UTC, with moment magnitude $4.9$ and with epicenter $136$ $km$ from Santiago
de Chile (see Table \ref{tab:detected}). Each bar in the graph of Figure \ref{fig:detection}
represents the value of $S\left( \varepsilon ,t\right) $ evaluated at the
arrival of a vibration signal. Before the earthquake, the arrivals are
sparse in time and $S\left( \varepsilon ,t\right) $ is low. When the
subnetwork senses the earthquake, the arrivals are close in time and $%
S\left( \varepsilon ,t\right) $ exceeds the threshold $h$ after few seconds.
The warning was issued $43$ seconds after the earthquake, though this delay
includes the time required to the earthquake waves to cover the distance
between the epicenter and Santiago de Chile. 

\subsection{Warning time}
\begin{figure}
  \includegraphics[width=0.48\textwidth]{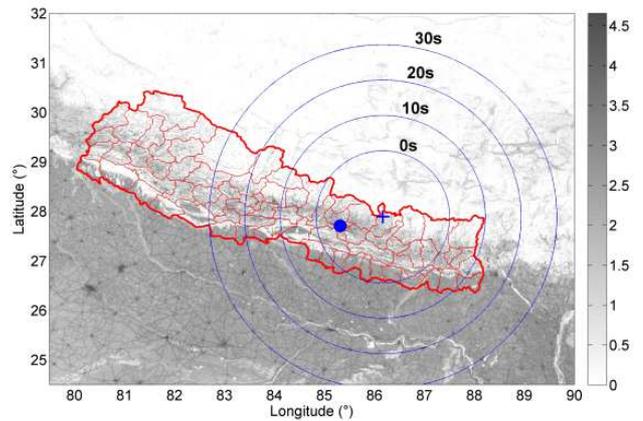}
\caption{Forewarning times for the magnitude $7.3$ earthquake of May 12, 2015 in Nepal. Legend: cross: earthquake epicenter; filled dot: Kathmandu subnetwork; circles: forewarning radius of $0$, $10$, $20$ and $30$ seconds; colormap: log$_{10}$-population.}
\label{fig:nepal}     
\end{figure}

To be effective, an EEW system should alert the population (or part of it) many seconds in advance in order to allow people to take cover. 
The image of Figure \ref{fig:nepal} depicts the warning times related to the magnitude $7.3$ earthquake event of May 12, 2015 with epicenter in Nepal. More than $150$ people were killed by the earthquake and more than $3,200$ people were injured.

The earthquake was detected by the Kathmandu subnetwork with a delay of $4$ $s$ from the first vibration signal received by the server. It is assumed here that the first smartphone detected the earthquake $1.5$ seconds from the beginning of the ground shaking and that the transmission delay from the smartphone to the server was $0.5$ $s$. Moreover, it is assumed that the server took $0.5$ $s$ to notify the users with the Earthquake Network application installed, bringing the total delay to $6.5$ $s$.
From the scientific data of the EMSC catalog, the speed of the earthquake waves is estimated to be $0.0715$ $^{\circ}/s$ and it is assumed to be isotropic.
Each circle in the image of Figure \ref{fig:nepal} represents a warning time. The population inside the $0$ $s$ circle received the earthquake warning after having experienced the earthquake while the population outside received the warning in advance, with a warning time proportional to the distance between each person and the epicenter. Figure \ref{fig:nepal} suggests that the EEW system is more effective when the earthquake is detected near the epicenter, in which case it is possible to notify in advance a higher population fraction.

\section{Detection delay simulation\label{sec:delay_simulation}}

A set of realistic simulations is carried out to understand the performances of the
Earthquake Network EEW system in terms of detection probability and
detection delay. 
The focus is on the network behavior under a seismic event,
rather than the physical dynamics of the seismic event itself. In particular it is implicitly assumed that the seismic intensity is constant over the area considered.

The main simulation parameter are the report fraction $\phi$ and the report spread $\sigma$. Indeed, when an earthquake is assumed to hit the geographic area of the subnetwork under study, $\phi$ is the fraction of active smartphones that reports a vibration signal. Since strong earthquakes are felt by a large fraction of smartphones, $\phi$ essentially describes the earthquake intensity. On the other side, $\sigma$ is the time distance between the first and the last vibration signals received by the server. It describes the (random) delays that prevent the vibration signals to be notified to the server instantaneously.

Simulations are implemented considering the data of the Santiago de Chile subnetwork, and the time frame of Table \ref{tab:subnetworks} with the real earthquake events filtered out. 
For each combination of report fraction and report spread
of Table \ref{tab:detection_fraction}, $1,000$ earthquakes are simulated as follows. 
Firstly, an earthquake time $\tau$ is randomly extracted over the range of the time frame. Secondly, the corresponding number of active smartphones $\nu_{\tau}$ is used to simulate $\nu_{\tau} \times \phi$ vibration signals, randomly located in the interval $(\tau,\tau+\sigma)$.
The detector $S(\epsilon,t)$ is then applied and, in the case of detection, the detection delay $t^*-\tau$ is evaluated, with $t^*$ the detection time.

Table \ref{tab:detection_fraction}
shows the detection fraction which is an estimate of the detection
probability. It is possible to note that the detection fraction is
influenced by the report fraction while the report spread has no effect. In
particular, when $\phi>0.25$, the earthquake is detected with about $%
90\%$ probability while $\phi>0.55$ guarantees almost sure detection.

\begin{table}[tbp]
\caption{Detection fraction (in percentage) of the detector $S(\epsilon,t)$, $\epsilon=30$ $s$, for $1,000$ simulated earthquakes with respect
to the report fraction $\phi$ and the report spread $\sigma$ (in seconds).}
\label{tab:detection_fraction}
\begin{tabular}{rr|ccccccc}
\hline
&  & \multicolumn{7}{|c}{$\sigma$} \\ 
&  & $2$ & $3$ & $5$ & $10$ & $15$ & $20$ & $25$ \\ \hline
\multicolumn{1}{c}{} & $0.01$ & \multicolumn{1}{|r}{$0.0$} & 
\multicolumn{1}{r}{$0.0$} & \multicolumn{1}{r}{$0.0$} & \multicolumn{1}{r}{$%
0.0$} & \multicolumn{1}{r}{$0.0$} & \multicolumn{1}{r}{$0.0$} & 
\multicolumn{1}{r}{$0.0$} \\ 
\multicolumn{1}{c}{} & $0.05$ & \multicolumn{1}{|r}{$3.4$} & 
\multicolumn{1}{r}{$4.3$} & \multicolumn{1}{r}{$4.6$} & \multicolumn{1}{r}{$%
5.1$} & \multicolumn{1}{r}{$6.2$} & \multicolumn{1}{r}{$4.8$} & 
\multicolumn{1}{r}{$5.1$} \\ 
\multicolumn{1}{c}{} & $0.10$ & \multicolumn{1}{|r}{$34.8$} & 
\multicolumn{1}{r}{$39.3$} & \multicolumn{1}{r}{$40.2$} & \multicolumn{1}{r}{%
$41.6$} & \multicolumn{1}{r}{$39.3$} & \multicolumn{1}{r}{$40.3$} & 
\multicolumn{1}{r}{$41.6$} \\ 
\multicolumn{1}{c}{} & $0.15$ & \multicolumn{1}{|r}{$67.3$} & 
\multicolumn{1}{r}{$67.1$} & \multicolumn{1}{r}{$68.5$} & \multicolumn{1}{r}{%
$68.2$} & \multicolumn{1}{r}{$69.4$} & \multicolumn{1}{r}{$68.1$} & 
\multicolumn{1}{r}{$68.6$} \\ 
\multicolumn{1}{c}{} & $0.20$ & \multicolumn{1}{|r}{$80.1$} & 
\multicolumn{1}{r}{$78.9$} & \multicolumn{1}{r}{$81.6$} & \multicolumn{1}{r}{%
$83.5$} & \multicolumn{1}{r}{$81.8$} & \multicolumn{1}{r}{$82.6$} & 
\multicolumn{1}{r}{$82.1$} \\ 
\multicolumn{1}{c}{} & $0.25$ & \multicolumn{1}{|r}{$90.9$} & 
\multicolumn{1}{r}{$90.8$} & \multicolumn{1}{r}{$90.4$} & \multicolumn{1}{r}{%
$90.1$} & \multicolumn{1}{r}{$90.2$} & \multicolumn{1}{r}{$90.5$} & 
\multicolumn{1}{r}{$89.6$} \\ 
\multicolumn{1}{c}{} & $0.30$ & \multicolumn{1}{|r}{$94.1$} & 
\multicolumn{1}{r}{$94.5$} & \multicolumn{1}{r}{$95.1$} & \multicolumn{1}{r}{%
$95.3$} & \multicolumn{1}{r}{$93.1$} & \multicolumn{1}{r}{$94.2$} & 
\multicolumn{1}{r}{$93.1$} \\ 
\multicolumn{1}{c}{} & $0.35$ & \multicolumn{1}{|r}{$96.3$} & 
\multicolumn{1}{r}{$97.2$} & \multicolumn{1}{r}{$97.1$} & \multicolumn{1}{r}{%
$96.7$} & \multicolumn{1}{r}{$97.0$} & \multicolumn{1}{r}{$97.0$} & 
\multicolumn{1}{r}{$97.1$} \\ 
\multicolumn{1}{c}{$\phi$} & $0.40$ & \multicolumn{1}{|r}{$99.2$} & 
\multicolumn{1}{r}{$98.7$} & \multicolumn{1}{r}{$98.6$} & \multicolumn{1}{r}{%
$98.5$} & \multicolumn{1}{r}{$98.4$} & \multicolumn{1}{r}{$99.1$} & 
\multicolumn{1}{r}{$98.8$} \\ 
\multicolumn{1}{c}{} & $0.45$ & \multicolumn{1}{|r}{$99.6$} & 
\multicolumn{1}{r}{$99.6$} & \multicolumn{1}{r}{$99.9$} & \multicolumn{1}{r}{%
$99.6$} & \multicolumn{1}{r}{$99.6$} & \multicolumn{1}{r}{$99.8$} & 
\multicolumn{1}{r}{$99.1$} \\ 
\multicolumn{1}{c}{} & $0.50$ & \multicolumn{1}{|r}{$99.6$} & 
\multicolumn{1}{r}{$99.9$} & \multicolumn{1}{r}{$100.0$} & 
\multicolumn{1}{r}{$100.0$} & \multicolumn{1}{r}{$99.9$} & 
\multicolumn{1}{r}{$99.9$} & \multicolumn{1}{r}{$99.8$} \\ 
\multicolumn{1}{c}{} & $0.55$ & \multicolumn{1}{|r}{$100.0$} & 
\multicolumn{1}{r}{$100.0$} & \multicolumn{1}{r}{$100.0$} & 
\multicolumn{1}{r}{$100.0$} & \multicolumn{1}{r}{$99.9$} & 
\multicolumn{1}{r}{$100.0$} & \multicolumn{1}{r}{$100.0$} \\ 
\multicolumn{1}{c}{} & $0.60$ & \multicolumn{1}{|r}{$100.0$} & 
\multicolumn{1}{r}{$100.0$} & \multicolumn{1}{r}{$100.0$} & 
\multicolumn{1}{r}{$100.0$} & \multicolumn{1}{r}{$100.0$} & 
\multicolumn{1}{r}{$100.0$} & \multicolumn{1}{r}{$100.0$} \\ 
\multicolumn{1}{c}{} & $0.65$ & \multicolumn{1}{|r}{$100.0$} & 
\multicolumn{1}{r}{$100.0$} & \multicolumn{1}{r}{$100.0$} & 
\multicolumn{1}{r}{$100.0$} & \multicolumn{1}{r}{$100.0$} & 
\multicolumn{1}{r}{$100.0$} & \multicolumn{1}{r}{$100.0$} \\ 
\multicolumn{1}{c}{} & $0.70$ & \multicolumn{1}{|r}{$100.0$} & 
\multicolumn{1}{r}{$100.0$} & \multicolumn{1}{r}{$100.0$} & 
\multicolumn{1}{r}{$100.0$} & \multicolumn{1}{r}{$100.0$} & 
\multicolumn{1}{r}{$100.0$} & \multicolumn{1}{r}{$100.0$} \\ 
\multicolumn{1}{c}{} & $0.75$ & \multicolumn{1}{|r}{$100.0$} & 
\multicolumn{1}{r}{$100.0$} & \multicolumn{1}{r}{$100.0$} & 
\multicolumn{1}{r}{$100.0$} & \multicolumn{1}{r}{$100.0$} & 
\multicolumn{1}{r}{$100.0$} & \multicolumn{1}{r}{$100.0$} \\ 
\multicolumn{1}{c}{} & $0.80$ & \multicolumn{1}{|r}{$100.0$} & 
\multicolumn{1}{r}{$100.0$} & \multicolumn{1}{r}{$100.0$} & 
\multicolumn{1}{r}{$100.0$} & \multicolumn{1}{r}{$100.0$} & 
\multicolumn{1}{r}{$100.0$} & \multicolumn{1}{r}{$100.0$} \\ \hline
\end{tabular}%
\end{table}

Table \ref{tab:detection_delay} reports, instead, the average detection
delay which is influenced by both $\phi$ and $\sigma$. As expected, the lower $\phi$ the lower the detection delay, while a lower $\phi$ implies a higher delay.

\begin{table}[tbp]
\caption{Average detection delay (in seconds) of the detector $S(\epsilon,t)$, $\epsilon=30$ $s$, for $1,000$ simulated earthquakes with respect
to the report fraction $\phi$ and the report spread $\sigma$ (in seconds).}
\label{tab:detection_delay}
\begin{tabular}{rr|rrrrrrr}
\hline
&  & \multicolumn{7}{|c}{$\sigma$} \\ 
&  & $2$ & $3$ & $5$ & $10$ & $15$ & $20$ & $25$ \\ \hline
& $0.01$ & \multicolumn{1}{|c}{$-$} & \multicolumn{1}{c}{$-$} & 
\multicolumn{1}{c}{$-$} & \multicolumn{1}{c}{$-$} & \multicolumn{1}{c}{$-$}
& \multicolumn{1}{c}{$-$} & \multicolumn{1}{c}{$-$} \\ 
& $0.05$ & $1.78$ & $3.08$ & $4.96$ & $11.18$ & $16.91$ & $19.32$ & $24.04$
\\ 
& $0.10$ & $1.38$ & $2.14$ & $3.48$ & $7.26$ & $11.12$ & $14.18$ & $17.50$
\\ 
& $0.15$ & $1.24$ & $1.85$ & $3.20$ & $6.61$ & $9.90$ & $12.61$ & $15.73$ \\ 
& $0.20$ & $1.09$ & $1.62$ & $2.79$ & $5.57$ & $8.38$ & $10.86$ & $13.59$ \\ 
& $0.25$ & $0.98$ & $1.44$ & $2.35$ & $4.88$ & $7.09$ & $9.56$ & $11.77$ \\ 
& $0.30$ & $0.87$ & $1.29$ & $2.14$ & $4.19$ & $6.42$ & $8.49$ & $10.53$ \\ 
& $0.35$ & $0.76$ & $1.13$ & $1.96$ & $3.84$ & $5.74$ & $7.64$ & $9.61$ \\ 
$\phi$ & $0.40$ & $0.68$ & $1.02$ & $1.74$ & $3.52$ & $4.99$ & $6.84$ & $8.59$
\\ 
& $0.45$ & $0.61$ & $0.91$ & $1.58$ & $3.17$ & $4.48$ & $6.25$ & $7.69$ \\ 
& $0.50$ & $0.57$ & $0.88$ & $1.40$ & $2.88$ & $4.27$ & $5.65$ & $7.14$ \\ 
& $0.55$ & $0.53$ & $0.78$ & $1.30$ & $2.53$ & $3.92$ & $5.01$ & $6.44$ \\ 
& $0.60$ & $0.49$ & $0.70$ & $1.19$ & $2.42$ & $3.52$ & $4.81$ & $5.85$ \\ 
& $0.65$ & $0.45$ & $0.67$ & $1.13$ & $2.22$ & $3.25$ & $4.35$ & $5.58$ \\ 
& $0.70$ & $0.41$ & $0.61$ & $1.06$ & $2.13$ & $3.07$ & $3.99$ & $5.15$ \\ 
& $0.75$ & $0.39$ & $0.58$ & $0.92$ & $1.97$ & $2.93$ & $3.87$ & $4.73$ \\ 
& $0.80$ & $0.37$ & $0.55$ & $0.87$ & $1.82$ & $2.72$ & $3.55$ & $4.52$ \\ 
\hline
\end{tabular}
\end{table}

\begin{figure}
  \includegraphics[width=0.48\textwidth]{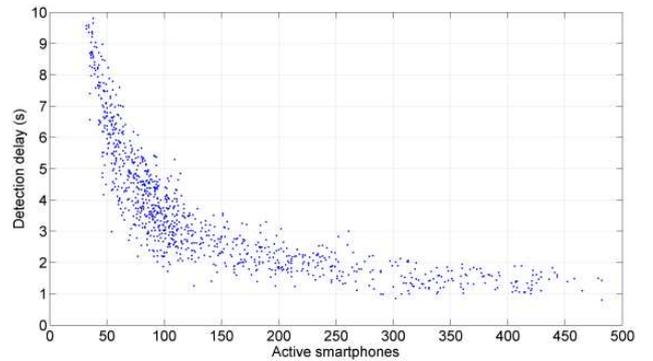}
\caption{Detection delay with respect to the number of active smartphones $\nu_{\tau}$. Dots represent the
detection delay for the $1,000$ simulated earthquakes with report fraction $\phi=0.5$ and report spread $\sigma=10$ $s$.}
\label{fig:active_vs_delay}     
\end{figure}

Note that the delays reported in Table \ref{tab:detection_delay} are averages
over $1,000$ simulations. The delay related to a specific earthquake is also
a function of the number of active smartphones. For instance, the graph of Figure \ref{fig:active_vs_delay}
shows the detection delays for $1,000$ simulations with $\phi=0.5$ and $\sigma=10$ $s$. The detection delay
can be as high as $10$ $s$ when the number of active smartphones is lower than $50$, while a number
of active smartphones higher than $300$ implies a delay between one and two seconds.

\subsection{Window size}
As mentioned above, the choice of the window size $\epsilon$ has an impact on the behavior of the detector $S(\epsilon,t)$.
The choice of the window size has to take into account that, for smaller $\epsilon$ a shorter delay is expected but, if the number of active nodes $\nu_t$ and/or the report fraction $\phi$ are low, then a small $\epsilon$ entails a low detection fraction. That said, the false alarm probability $\alpha$ is controlled for each $\epsilon$ since the threshold $h$ is computed for fixed $\epsilon$.

\begin{figure}
  \includegraphics[width=0.48\textwidth]{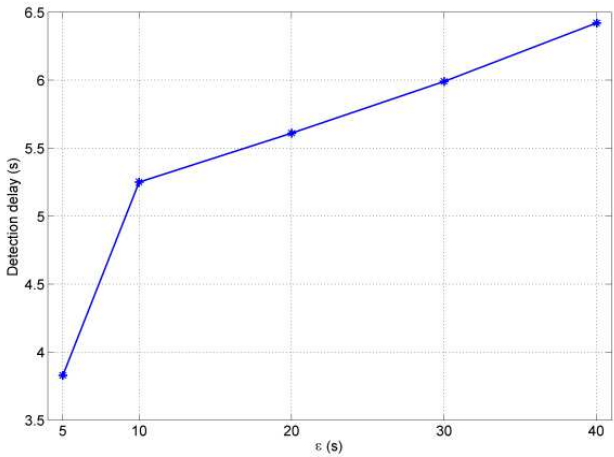}
\caption{Detection delay vs window size $\epsilon$. Delays are averaged over $\phi$ and $\sigma$ defined in Table \ref{tab:detection_fraction}.}
\label{fig:DetectDelay_vs_Eps}     
\end{figure}

\begin{figure}
  \includegraphics[width=0.48\textwidth]{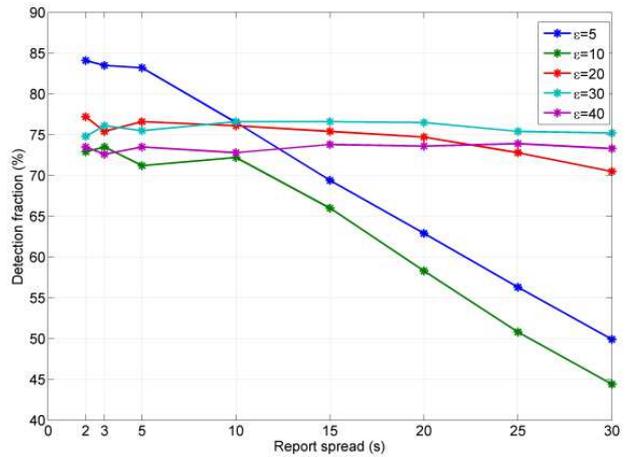}
\caption{Detection fraction vs report spread $\sigma$ for various window sizes $\epsilon$. Detection fractions are averaged over $\phi$ defined in Table \ref{tab:detection_fraction}.}
\label{fig:DetectFrac_vs_RepLen}     
\end{figure}

The simulations and statistics as in Tables \ref{tab:detection_fraction} and \ref{tab:detection_delay} have been performed for $\epsilon=5,10,20,30,40$ $s$.
In Figure \ref{fig:DetectDelay_vs_Eps}, the detection delay, averaged for $\phi$ and $\sigma$, shows that increasing window size slows down the earthquake detection. On the other side, Figure \ref{fig:DetectFrac_vs_RepLen} shows that for small $\epsilon$ the detection fraction is poor when the spread $\sigma$ is large.
For these reasons, $\epsilon=30$ $s$ used in this work is a good compromise between early detection, which applies to short spread events, and acceptable detection fraction for large spread.

\section{Conclusions\label{sec:conclusions}}

EEW systems may save human lives in the case of destructive earthquake events. Especially in underdeveloped and developing countries,
however, the proliferation of EEW systems may be dampened by the high installation and operating costs. In this paper it is shown that
crowdsourced EEW systems based on smartphones can be used to detect earthquakes in quasi real-time and to alert the population through
the very same devices, with zero installation costs and very low operating costs.

The problem of detecting earthquakes from the data sent by the smartphone network has been solved through a statistical approach which is able to handle a dynamic network and, more importantly, which allows to control the probability of false alarms.
The detection capabilities of the approach have been proven using real data collected by the crowdsourced EEW system of the Earthquake Network project. Considering three subnetworks of the smartphone network, the system was able to detect earthquakes down to magnitude $4$ and with detection delays ranging from $2$ to $17$ seconds. The variability of the detection delay is quite high within and across subnetworks and it might be reduced considering the information on the spatial location of the smartphones. Indeed, the statistical approach developed in this paper does not fully exploit this information, which is only used to discriminate across subnetworks. Nonetheless, the approach has the advantage of being computationally fast and thus suitable for quasi real-time detection. Future works will try to extend the statistical approach in order to use the information on the smartphone location. This will allow to introduce a global detector which is not based on subnetworks and which is able to provide an estimate of the earthquake epicenter.

%\begin{acknowledgements}
%\end{acknowledgements}

%\bibliographystyle{plain}      % basic style, author-year citations
\bibliographystyle{plain}       % APS-like style for physics
\bibliography{biblio}   % name your BibTeX data base

\end{document}